# Deriving Priors for Bayesian Prediction of Daily Response Propensity in Responsive Survey Design: Historical Data Analysis vs. Literature Review

**Running Header:** Deriving Priors for Bayesian Prediction of Response Propensity

**Brady T. West (Corresponding Author)**[1]
Survey Research Center
Institute for Social Research
University of Michigan-Ann Arbor
426 Thompson Street
Ann Arbor, MI, USA, 48106-1248
Phone: 734-647-4615

**James Wagner**
Survey Research Center
Institute for Social Research
University of Michigan-Ann Arbor

**Stephanie Coffey**
U.S. Census Bureau, and
Joint Program in Survey Methodology

**Michael R. Elliott**
Survey Research Center
Institute for Social Research, and
Department of Biostatistics
University of Michigan-Ann Arbor

**Abstract Word Count:** 219

**Manuscript Word Count:** 7,224

[1] This work was supported by a grant from the National Institutes for Health (#1R01AG058599-01). The 2010-2020 National Survey of Family Growth (NSFG) was conducted by the Centers for Disease Control and Prevention's (CDC's) National Center for Health Statistics (NCHS), under contract # 200-2010-33976 with University of Michigan's Institute for Social Research with funding from several agencies of the U.S. Department of Health and Human Services, including CDC/NCHS, the National Institute of Child Health and Human Development (NICHD), the Office of Population Affairs (OPA), and others listed on the NSFG webpage (see http://www.cdc.gov/nchs/nsfg/). The data collection included in this paper was performed under contract # 200-2010-33976. The views expressed here do not represent those of NCHS nor the other funding agencies. Neither the study design nor the analysis presented were preregistered.

**Abstract**

Responsive Survey Design (RSD) aims to increase the efficiency of survey data collection via live monitoring of paradata and the introduction of protocol changes when survey errors and increased costs seem imminent. Daily predictions of response propensity for all active sampled cases are among the most important quantities for live monitoring of data collection outcomes, making sound predictions of these propensities essential for the success of RSD. Because it relies on real-time updates of prior beliefs about key design quantities, such as predicted response propensities, RSD stands to benefit from Bayesian approaches. However, empirical evidence of the merits of these approaches is lacking in the literature, and the derivation of informative prior distributions is required for these approaches to be effective. In this paper, we evaluate the ability of two approaches to deriving prior distributions for the coefficients defining daily response propensity models to improve predictions of daily response propensity in a real data collection employing RSD. The first approach involves analyses of historical data from the same survey, and the second approach involves literature review. We find that Bayesian methods based on these two approaches result in higher-quality predictions of response propensity than more standard approaches ignoring prior information. This is especially true during the early-to-middle periods of data collection, when survey managers using RSD often consider interventions.

## 1. Introduction

In an effort to minimize survey errors and costs, survey methodologists have developed a conceptual framework for survey data collection known as responsive survey design (RSD; Groves and Heeringa, 2006). RSD monitors the quality and cost efficiency of a survey data collection in real time, enabling informed decisions about design changes in response to the large uncertainties that accompany survey research. Unfortunately, current implementations of RSD are often ad hoc and simplistic, failing to integrate prior knowledge of data collection outcomes with incoming real-time information. Because survey data collection is fraught with uncertainty, survey researchers employing RSD need to compute sound predictions of unknown design quantities (e.g., subgroup response rates) to improve the efficiency of their survey designs.

Schouten et al. (2018) recently proposed a general Bayesian framework for improving the accuracy of predictions of survey design quantities across all stages of a survey design. Following this general framework, key survey design quantities (e.g., the probability that a household contains an eligible person, costs per sampled unit, unit-level response propensities, etc.) are treated as *random variables*, and parametric models are specified for these random variables. The parameters defining these models are then assigned prior distributions, and these prior distributions are then updated given the new information obtained from a current data collection. Predictions of the parameters are then drawn repeatedly from their resulting posterior distributions, and subsequent draws of the design quantities of interest can be computed based on the specified models. These types of Bayesian updating methods have recently gained popularity in economic forecasting and business management contexts as well (McCann and Schwab, forthcoming).

Given these simulated draws of the design quantity of interest, one can then make real-time design decisions that reflect uncertainty about the survey process and leverage useful prior information. Ideally, these design decisions would be embedded in a larger RSD framework with a specified loss function. With this paper, our goal is to improve predictions of a design quantity that plays a crucial role in such a decision framework: the probability that a given sampled unit will respond at the next call attempt. We attempt to improve predictions of these *daily response propensities* by focusing on alternative approaches to deriving informative prior distributions for the regression coefficients used to compute them, given how important useful prior information is in the Schouten et al. (2018) framework. We do not aim to develop approaches for modifying a survey data collection protocol given these improved predictions, but rather to examine whether particular sources of prior information produce better predictions in real-time. We revisit the issue of modifying designs in light of the results in this paper in the Discussion.

To date, several responsive and adaptive designs have used predicted response propensities as inputs (e.g., Rosen et al., 2014), where response propensity is assumed to assess the "quality" of the active sample (Groves and Heeringa, 2006; Groves et al., 2009; West and Groves, 2013). Other surveys have used response propensity models to make decisions about design features (Peytchev et al., 2009; Wagner, 2013; Jackson et al., 2019). For example, Wagner (2013) used estimated contact propensities to trigger decisions about the timing of the next call for each active case. The models producing these estimates used the data available at each point in the data collection process. However, early in the field period, there may be a relatively high prevalence of "easy" early responders, and estimates of response propensity may be both biased and noisy due to the limited accumulation of helpful paradata (Wagner and Hubbard, 2014). Early responders may also be different from late responders in ways that are not observable early in the period, and in face-to-face surveys, interviewers may select cases to attempt based on features not available in the paradata (Kennickell, 2003). Any of these situations may lead to estimated response propensity models that generate inaccurate predictions. These inaccuracies can reduce the efficiency of RSD by suggesting design modifications at less-than-optimal points in time.

Tourangeau et al. (2017) recently reviewed a number of studies that presented mixed evidence of the effectiveness of RSD. Among possible explanations for this mixed evidence were the generally difficult climate in which surveys are currently conducted, the high costs involved with varying powerful design features (e.g. high incentives or new modes), and inefficient designs. Imprecise timing of design modifications during data collection could be one source of this inefficiency, which can lead to higher costs or mitigate bias reduction. Unstable predictions of response propensity that feed into these modifications early in a fieldwork period can lead to a misallocation of field effort that increases bias rather than reducing it. Similarly, targeting cases solely based on predicted response propensities, when those predictions are inaccurate, may not necessarily allocate effort toward bias reduction. Leveraging prior information about the parameters used to compute the predictions of response propensity has the potential to reduce some of these inefficiencies in data collection that may be limiting RSD from reaching its full potential. However, where should survey managers search for useful prior information about these parameters? This is a key question that we seek to answer with the current study.

Schouten and colleagues (2018) presented a simulation study evaluating the ability of their general Bayesian framework to improve predictions of costs, contact propensities, and cooperation propensities in a survey employing adaptive design principles by leveraging prior information about the parameters defining these predictive models. Schouten et al. find that Bayesian approaches can provide benefits, *provided that the priors are informative*. This underscores the importance of sound choices of prior distributions for this type of approach to be fully effective in improving the efficiency of survey designs. Schouten et al. call for analyses examining the sensitivity of predictions based on this general approach to alternative choices of prior distributions. Given the general lack of empirical work on this topic in the literature, we aim to evaluate alternative approaches to deriving prior information and their ability to improve predictions of daily response propensities *during* data collection.

Figure 1 presents an overview of our general approach. We focus on two approaches to deriving prior distributions for response propensity model coefficients in an RSD framework (highlighted in boldface in Figure 1): analyzing historical data sets from similar surveys and intensive literature review. Specifying priors for the parameters that define key survey design quantities, specifically the coefficients of response propensity models in our study, is no simple task to carry out. Of particular interest are parameters for which there are not pre-existing data. In this case, prior knowledge does not exist in the form of historical data from a previous study, and may only exist in the form of published literature or expert opinion. We do not consider the derivation of priors based on expert opinion (Coffey et al., 2020; see Figure 1) in our evaluation; we revisit this idea in the Discussion. We simultaneously evaluate the prior distributions derived from analyses of historical data and intensive literature review with respect to their ability to improve predictions of response propensity at different points in time for a single real survey, relative to more "standard" (or non-informative) approaches that ignore prior information and only leverage information from the current data collection.

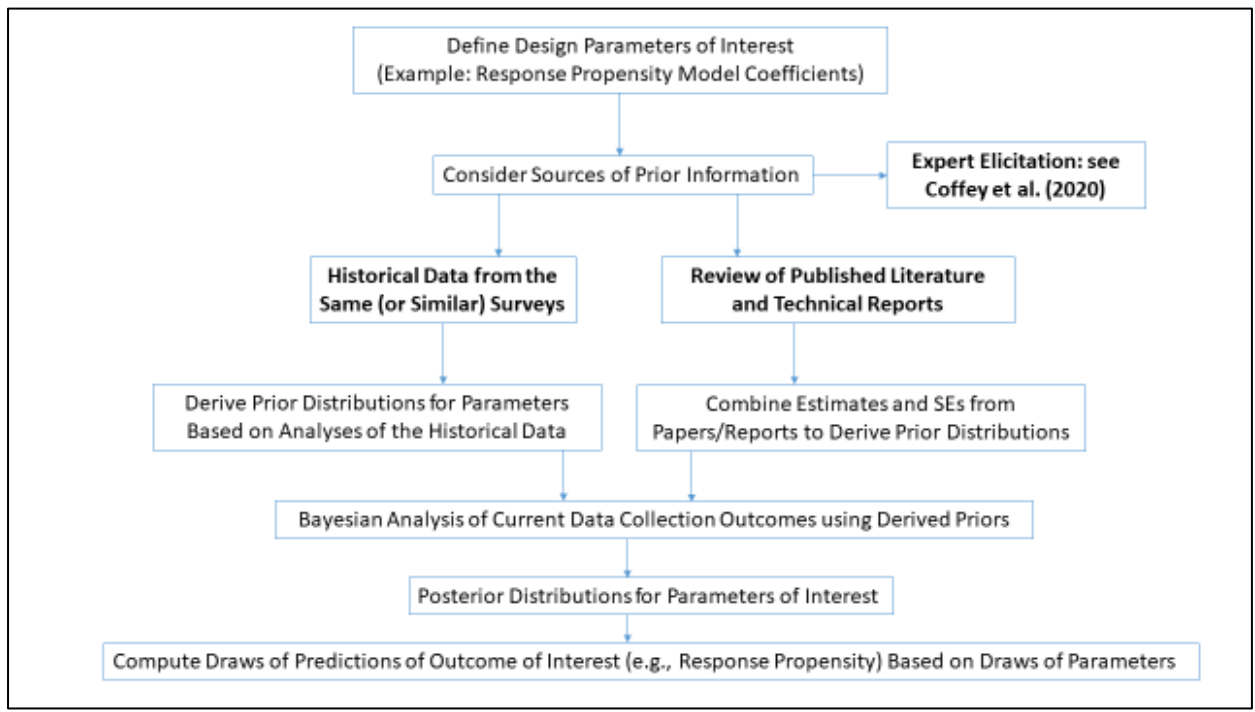

**Figure 1.** Overview of Alternative Approaches to Deriving Prior Distributions and Computing Posterior Predictions of Daily Response Propensity in our Study (SEs = standard errors).

We consider prediction of daily response propensities in the National Survey of Family Growth (NSFG), a face-to-face cross-sectional survey of U.S. households containing at least one person aged 15-49 that is repeated every calendar quarter. This is an ideal test bed for Bayesian RSD, since we can use the data from previous quarters to derive prior distributions for response propensity model coefficients. We can then compare these priors with priors developed using reviews of the literature, using the observed responses from a target quarter of data collection to measure the quality of the predictions.

## 2. Methods

*2.1 Overview of the NSFG RSD*

The NSFG selects a national sample of U.S. housing unit addresses each quarter of the year, and attempts to collect fertility and family formation data from randomly selected persons living at the sampled addresses. The target population from which the NSFG selects these four independent national samples is persons living in the U.S. who are between the ages of 15 and 49. Interviewers first visit randomly sampled households and attempt to screen the households for eligible persons based on their age. After completing the screening interview and confirming that eligible persons are present in the household, the interviewers use a computer algorithm to randomly select one of the eligible individuals to participate in the main survey interview. This interview, which the selected person can either schedule for a future date or complete immediately following the screening interview, usually takes 60-80 minutes and covers a variety of fertility-related topics. Additional details can be found in Lepkowski et al. (2013).

NSFG managers analyze paradata on a daily basis, modeling the probability that active sample households will respond to either the screening interview or the main interview at the next contact attempt. The managers might use these predictions for prioritization of active cases (e.g., Wagner et al., 2012) or when selecting a subsample of active cases for a new "second phase" data collection protocol that begins after 10 weeks of data collection in every data collection quarter. In selecting this sample, managers may over-sample high-propensity cases to receive the increased effort from interviewers and the offers of higher incentives that define the second phase protocol. Alternatively, managers may choose to over-sample low-propensity cases to receive the increased effort if lower-propensity cases are also predicted to have unique values on the variables of interest (in an effort to reduce nonresponse bias). Accurate model-based predictions are thus essential for maximizing the efficiency of the data collection effort in any given quarter. For purposes of this study, we focus on models for the probability of responding to the initial *screening* interview.

*2.2 Modeling Screening Response Propensity in the NSFG*

For this study, we analyzed data from 13 quarters of the NSFG (roughly ranging in time from June 2013 to September 2016). Across these quarters, screening interview response rates tended to be very stable, and associations of the various covariates that we considered in this study with response propensity tended to remain stable as well. We sought to evaluate predictions of the probability that individuals in *each of the five most recent quarters* (i.e., June 2015 to September 2016) will respond to a screening interview at a given contact attempt, considering the *eight preceding quarters* in each case as historical data that may be useful for defining prior distributions for the response propensity model coefficients. For example, for the data collection quarter from June 2016 to September 2016 (Quarter 20 from the 2011-2019 NSFG), we considered historical data from Quarters 12-19 for our analyses.

Although model selection is not the focus of the current paper, we needed to select a model. We initially fitted a discrete-time logit regression model to a stacked data set containing the outcomes of all screening interview attempts during the eight most recent quarters (Quarters 13 through 20). In this model, the dependent variable was a binary indicator of successful completion of a screening interview at that contact attempt. The candidate predictor variables included NSFG paradata, sampling frame information, and linked commercial data, each of which have been employed for the prediction of response propensity in prior studies using NSFG data (West, 2013; West and Groves, 2013; West et al., 2015). We employed a backward elimination approach to identify a common set of significant predictors of screener response propensity at a given contact attempt. After identifying the significant predictors ($p < 0.05$, based on Wald tests), we manually added two predictor variables that were deemed important to monitor by NSFG managers: type of sampling area (non-self-representing units, larger self-representing units, and the three largest MSAs) and socio-demographic domain of the sampled area segment based on U.S. Census data (<10% Black, <10% Hispanic; >10% Black, <10% Hispanic; <10% Black, >10% Hispanic; >10% Black, >10% Hispanic).

Table A1 in the online appendix describes the set of predictors that we identified as significant after applying the backward elimination procedure to the contact attempt data from the eight most recent quarters. These significant predictor variables, which resulted in 72 coefficients in the model, included:

1. The fielding day of the quarter (1 through 84) (Recall that days from the last two weeks of the quarter [71 through 84] featured the "second phase" data collection protocol of the NSFG RSD, which was designed to increase response propensity for difficult cases with lower response propensity in general);
2. Paradata (e.g., number of contact attempts made, expressions of resistance at previous contact attempts, and number of contacts established up to that day);
3. Sampling frame information (e.g., type of sampling area); and

4. Linked commercial variables (e.g., variables purchased from Marketing Systems Group, or MSG, such as the age of the head of the household).

Each of these variables could be used in theory to predict the probability of responding to the screening interview at a given contact attempt. One could also allow for non-linear relationships of fielding day with response propensity, or interact fielding day with the other predictor variables, if these types of non-linear or moderating relationships are expected or improve the fit of a model. There are implications of such modeling decisions for prior derivation, in that one would need to find historical estimates of these coefficients or prior studies using these more complex types of response propensity models. Based on a total of *n* = 119,981 contact attempts across these eight quarters, the Nagelkerke pseudo R-squared for our final fitted model was 0.09 (area under the receiver operating characteristic (ROC) curve (AUC) = 0.66).

*2.3 Overview of Approaches to Prior Derivation*

We aim to derive prior distributions for the coefficients in our daily response propensity models that stabilize and improve the accuracy of estimates of these coefficients, in turn generating better predictions of daily response propensity. We draw on two methods for deriving priors:

1) Historical estimates generated from prior data collection periods for the same survey; and

2) Detailed review of any literature presenting response propensity models that included similar covariates.

Following the general framework by Schouten et al. (2018), we define the normal prior distribution of the vector of coefficients for the predictor variables as

$$\boldsymbol{\beta} \sim N(\boldsymbol{\gamma}, \boldsymbol{\Gamma}), \quad (1)$$

where $\gamma$ is a mean of these coefficients obtained from one of the two sources of prior information introduced above, and $\Gamma$ is a variance-covariance matrix of these coefficients based on the same sources.

We note that this approach is strictly Bayesian, as we are assuming an informative prior distribution for these coefficients in the specific data collection wave of interest. We also note that the effect of this prior will diminish as more data from the current data collection quarter accumulate, since the posterior is the product of the (fixed) prior and the likelihood (which will increase as the incoming data from the current quarter increases). We now define these two approaches to deriving prior information about these coefficients in detail.

*Approach 1: Historical Data Analysis.* For this first approach, we fit a discrete-time hazard model of the form

$$\log\left(\frac{P(Y_{it}=1|\mathbf{X}_i)}{1-P(Y_{it}=1|\mathbf{X}_i)}\right) = \boldsymbol{\beta}^T \mathbf{X}_{it} = \sum_{c=0}^{C} \beta_c X_{itc} \qquad (2)$$

to the historical data using a standard (non-Bayesian) maximum likelihood method, with the purpose being to compute plausible values of the parameters defining the prior distribution. For a given set of historical data, $Y_{it}$ is an indicator of a completed screening interview at contact attempt $t$ for sampled person $i$ (where $Y_{it}=0$ for all unsuccessful contact attempts, and $Y_{it}=1$ for a completed screening interview, with a case not contributing any additional data after the screening interview is completed). The vector $\mathbf{X}_{it}$ consists of values on $c = 1, \ldots, C$ key predictors from those quarters (where $X_{it0} \equiv 1$ for the intercept). These can include regional factors, including local Census measures of aggregate demographics, as well as paradata measures, such as outcomes of previous call attempts and interviewer observations (see supplementary Table A1 for all predictors considered). The predictor variables may vary in their values across contact attempts indexed by $t$; examples include an indicator of whether contact attempt $t - 1$ resulted in a successful contact, and the number of contact attempts that had already been made for this case leading up to contact attempt $t$.

Next, given maximum likelihood estimates of the coefficients in (2) and their variances and covariances, obtained from one or multiple prior data collection periods, we consider three possible methods for defining the normal prior in (1):

1. A *standard* method that uses completely non-informative priors (i.e., $\boldsymbol{\gamma} = \mathbf{0}$ and $\boldsymbol{\Gamma} \approx diag(\infty)$) and ignores prior information when analyzing data from the current data collection quarter (mimicking what is often done in RSD).

2. A *precision-weighted prior* (PWP) method, where we first fit separate response propensity models of the form in (2) to the final accumulated contact attempt data from each of the eight prior NSFG quarters (indexed by $q$). The mean of the normal prior in (1) for the coefficient of predictor variable $c$ in quarter $q = 9$, for example, is then defined by

$$\boldsymbol{\gamma} = \left( \sum_{q=1}^{8} \text{var}(\hat{\boldsymbol{\beta}}_q)^{-1} \right)^{-1} \sum_{q=1}^{8} \text{var}(\hat{\boldsymbol{\beta}}_q)^{-1} \hat{\boldsymbol{\beta}}_q, \qquad (3)$$

where $\hat{\boldsymbol{\beta}}_q$ and $\text{var}(\hat{\boldsymbol{\beta}}_q)$ are, respectively, the estimate of $\boldsymbol{\beta}$ from (2) and its associated estimated variance-covariance matrix for quarter $q$, and the variance of the normal prior is defined by

$$\Gamma = \left(\sum_{q=1}^{8} \text{var}(\hat{\boldsymbol{\beta}}_q)^{-1} / 8\right)^{-1} = 8\left(\sum_{q=1}^{8} \text{var}(\hat{\boldsymbol{\beta}}_q)^{-1}\right)^{-1}. \qquad (4)$$

This formulates the prior using the eight previous quarters by weighting up quarters with more accurate estimates of $\boldsymbol{\beta}$, but setting the prior variance as the inverse of the mean of the prior precisions for each quarter; this approach is consistent with the concept of an informative *power prior* (Ibrahim et al., 2015). In order to stabilize $\text{var}(\hat{\boldsymbol{\beta}}_q)$, we used a ridge regression type estimator, replacing $\text{var}(\hat{\boldsymbol{\beta}}_q)$ with $(1-\lambda)\text{var}(\hat{\boldsymbol{\beta}}_q) + \lambda \, \text{diag}(\text{var}(\hat{\boldsymbol{\beta}}_q))$, where $\lambda = .003$ was chosen to ensure the invertibility of all values of $\text{var}(\hat{\boldsymbol{\beta}}_q)$.

When using the PWP method, one possible approach would be to apply more weight to coefficients from more recent quarters when defining the mean and variance of the normal prior in (1), or to drop prior data collection periods / seasons where response rates varied substantially, predictors of response propensity varied substantially, or completely different data collection protocols were employed. As we noted above, response rates to the NSFG screening interview, the coefficients in the response propensity models, and the NSFG data collection protocols were very stable across prior NSFG quarters, meaning that it would be reasonable to combine information from prior NSFG quarters without a need for weighting or the deletion of data from prior quarters. This may or may not be the case for other surveys, where one could apply more or less weight (including weights of zero) to the prior information, or even use a time series model to estimate parameters depending on the consistency of the data collections.

3. A *most recent period* (LAST) prior method, where the mean of the normal prior in (1) for the coefficients of the predictor variables in quarter $q$ is defined as

$$\gamma = \hat{\boldsymbol{\beta}}_{(q-1)}, \tag{5}$$

where $\hat{\boldsymbol{\beta}}_{(q-1)}$ is the vector of maximum likelihood estimates of the coefficients based on the final accumulated contact attempt data from the most recent quarter ($q - 1$), and the variance of the normal prior is defined as

$$\Gamma = \text{var}(\hat{\boldsymbol{\beta}}_{(q-1)}), \tag{6}$$

or the estimated variance-covariance matrix of the estimated coefficients from the most recent quarter. This approach can be thought of as a weighted combination of recent time periods where the most recent period receives all of the weight. In our analysis, we found that results based on the LAST method tended to be unstable across the five target quarters, which was likely due to unreliable estimation of some of the covariances of the estimated coefficients based on one quarter only. We therefore also considered a modified version of the LAST method, denoted by LASTZ, where all of the covariances in (6) were set to zero (i.e., we used independent normal priors for each of the coefficients based on the final fitted model from the most recent quarter).

We note that in general the most appropriate prior derivation method for historical data will depend strongly on trends and seasonality in response propensities. If response propensities are decreasing and/or changing significantly over time, then older historical data may well be less useful and should receive less weight, as we note above, and the LAST/LASTZ methods should provide improved performance. If there has been hardly any change in response rates over time and the coefficients of the response propensity model have remained relatively stable (which is the case with the NSFG), then using as much historical data as possible to inform the prior (e.g., the PWP method) should be beneficial.

*Approach 2: Literature Review.* For this approach, denoted by LIT moving forward, we reviewed the methodological and statistical literature to find empirical studies of survey response propensity as a function of predictors similar to those under consideration in the present study, at either the case or contact attempt levels. We considered peer-reviewed manuscripts, conference proceedings papers, official survey documentation, and technical reports. We used the following search terms in Google Scholar: "survey response propensity", "response propensity model", "survey response logistic", "survey response prediction", "covariates survey response", and "features survey response". We also visited the websites of the 13 statistical agencies in the U.S. (e.g., Census Bureau, National Center for Health Statistics, Bureau of Labor Statistics, etc.) to search for official documentation that included estimates of response propensity models. While Google Scholar provides researchers with a tool for searching the global literature on a topic, individuals interested in implementing this approach outside of the U.S. may need to work with statistical agencies directly to obtain information about response propensity models if these types of technical reports are not available online.

Once we identified a document, we first confirmed that the propensity discussed was in fact response propensity, rather than the propensity of another event (e.g., contact or refusal). Next, we checked whether the document included coefficients for the propensity model and their standard errors. Without this information, we would be unable to utilize the document for prior construction. Finally, we considered whether the target population of the survey was relatively broad, and likely to overlap with the NSFG. We excluded documents that discussed smaller or more specific target populations.

We identified 21 documents that discussed response propensity relevant to the NSFG population. We excluded 11 documents that did not provide coefficients and standard errors. Of the remaining 10 documents, we excluded one because it duplicated another document, and another due to its narrow target population. The eight studies that we retained included Olson and Groves (2009), Schonlau (2009), Peress (2010), Dahlhamer and Jans (2011), Hill and Shaw (2013), West and Groves (2013), Rosen et al. (2014), and Plewis and Shlomo (2017). We extracted the coefficients and standard errors reported in these studies, and we have provided these in the online supplementary materials.

We first established a crosswalk between the predictors analyzed in these eight studies and the NSFG predictors under evaluation in this study. For all but one of the studies, the reported coefficients and standard errors were on the log-odds scale. Peress (2010) reported results for a probit model, so prior to aggregating results across the studies, we multiplied the point estimates and standard errors from the probit model by 1.61, using the approximate result from Amemiya (1981). After performing these transformations, we determined the number of studies $S_c$ that reported coefficients for the same predictor variable $c$ used in the NSFG model. Then, the mean of the reported coefficients (on the log-odds scale) for that predictor $c$ was used as the mean of the normal prior distribution for that coefficient,

$$\gamma_c = \sum_{s=1}^{S_c} \hat{\beta}_{sc} / S_c, \qquad (7)$$

and the mean of the reported variances for the coefficients was used as the variance of the normal prior distribution for that coefficient:

$$\lambda_c = \sum_{s=1}^{S_c} \text{v\^ar}(\hat{\beta}_{sc}) / S_c. \tag{8}$$

We note that, in contrast to the use of historical data, the lack of overlap in predictors in the same model in a given paper means that it is difficult to estimate with any certainty the covariances between different regression parameter components. Hence, we set these covariances to 0, so that, in contrast to the historical data-derived priors (and similar to the LASTZ approach), $\Gamma = diag(\lambda_1,...,\lambda_C)$. We also note that while all of the articles that we identified for forming these priors are relatively new (no earlier than 2009), one could apply more weight to more recent articles when computing the mean and variance of each normal prior distribution based on (7) and (8). For example, one could potentially weight the contribution of each coefficient / variance estimate by the inverse of the number of five-year periods preceding the year for which the prior is being developed (e.g., articles from 2015-2019 would get a weight of 1, articles from 2010-2014 would get weight of 0.5, etc.). We tried this type of weighting scheme in the current analysis and found that our results using the literature review hardly changed. We note that this could be an artifact of the relatively stable response rates achieved over the time period under consideration.

Of the 72 coefficients identified as significant in our backward elimination approach described earlier (see Appendix Table A1), we were able to find prior information in the literature for 33 of them (46%). The means and variances of the normal prior distributions for the remaining 39 coefficients were set to 0 and 10, respectively, indicating a lack of information in the literature about these coefficients (i.e., we used nearly non-informative prior distributions when there was no evidence available in the literature).

*2.4 Analytic Approach*

*Step 1: Defining Benchmark Response Propensity Values.* We first fit a discrete time hazard model to the *final* accumulated contact attempt data for one of the five *target* quarters, using all available information from that quarter for the final set of predictors described earlier. We used the maximum likelihood estimates of the regression coefficients in this model, which had the form introduced earlier in (2), to compute a "final" predicted probability of response for each sampled person *i* at the *final* contact attempt made to that case, using all available information accumulated during that quarter:

$$\hat{p}_{i,final} = \frac{\exp(\hat{\boldsymbol{\beta}}^T \mathbf{X}_{i,final})}{1+\exp(\hat{\boldsymbol{\beta}}^T \mathbf{X}_{i,final})} \qquad (9)$$

These "final" predictions for each sampled person in (9) served as our benchmarks, quantifying the probability that a given person would respond at a contact attempt based on all of the information learned about that sampled person during the data collection (e.g., how many contact attempts were required, did a case last until the second phase, etc.). We chose these "final" predictions as targets because the ultimate goal of most RSDs is to reach an optimal cost/quality tradeoff by the *end* of data collection.

RSD attempts to introduce changes to the data collection protocol in hopes of introducing cost efficiency and reducing survey errors, thereby interfering with the "final" estimate of response propensity that would exist were no intervention carried out. Some of these changes may focus on increasing the probability that an active sampled person will cooperate with the survey request (e.g., offering a higher incentive amount during the second phase of the data collection quarter). Each NSFG quarter uses the same RSD approach, introducing the protocol change after 10 weeks of data collection, and possibly telling interviewers to prioritize certain groups of cases earlier in the first phase (Wagner et al., 2012). The "benchmark" predicted probabilities in (9) are based on all of the data gathered in a given quarter. They therefore reflect "best estimates" of the probability that a case will respond once they have been subjected to the *full* NSFG RSD in a given quarter. We choose these "best estimates" as the targets of our analysis, given the importance of predicted response propensity to the NSFG RSD (where, e.g., the stratification used for two-phase sampling depends on the predictions of response propensity). As we noted earlier, other parameters could certainly be targets of interest in other survey designs.

The consistency of the NSFG RSD across quarters is important when considering whether prior information regarding the coefficients for the model in (2) will in fact improve estimates of the coefficients in (2) earlier in a data collection period, and therefore the predictions of response propensity based on those coefficients. In other surveys, where different types of design modifications may be implemented in different data collection periods, "best estimates" of expected response propensity may need to leverage prior information that is obtained *only* from prior periods that employed *similar* types of designs. In this sense, we expect that the historical data approaches may tend to outperform predictions based on the literature review.

*Step 2: Compute Bayesian Predictions of Daily Response Propensity.* For each of the five most recent NSFG quarters (Quarters 16 through 20), we first generated the alternative prior distributions in (1) for the 72 response propensity model coefficients, using the derivation methods described above. For each alternative prior derivation approach, we then followed these steps:

1. Beginning on the seventh day of the current quarter (allowing for the accumulation of one week of information), where days are indexed by $d$ (hence, $d = 7$ to start), we used PROC MCMC in the SAS software (Version 9.4; 100 tuning steps, and 5,000 Monte Carlo simulations) to simulate posterior draws of the logistic regression model coefficients, given the *fixed* prior specifications in (1), the likelihood function for the logit model in (2), and the current cumulative data on that day $d$. We also considered a *dynamic* prior specification approach that updated the prior on each day (i.e., used the posterior for the regression coefficients on a given day as the new prior), but that approach did not yield substantively different results and was also more computationally intensive. We therefore used fixed priors for this approach.

2. For each draw $k$ of the coefficients on day $d$, we computed a corresponding draw of the predicted probability of responding to contact attempt $t$ for case $i$ on that day $d$:

$$\hat{p}_{itd}^{(k)} = \frac{\exp(\boldsymbol{\beta}_d^{(k)T}\mathbf{X}_{itd})}{1+\exp(\boldsymbol{\beta}_d^{(k)T}\mathbf{X}_{itd})}. \qquad (10)$$

We then averaged the 5,000 draws of the predicted probabilities for a given case $i$ receiving contact attempt $t$ on that day, and denoted this as the predicted probability of response for that case at that contact attempt on this day:

$$\hat{p}_{itd} = \sum_{k=1}^{5,000} \hat{p}_{itd}^{(k)} / 5,000 \qquad (11)$$

3. We then computed the difference between the predicted probability of response for case $i$ at attempt $t$ on day $d$ from (11) and the "final" predicted probability of response for case $i$ from (9):

$$diff_{itd} = \hat{p}_{itd} - \hat{p}_{i,final} \qquad (12)$$

4. Next, we computed the mean difference on day $d$ (as an estimate of bias, $B$) and the standard error of the mean difference, where $s_d$ is the standard deviation of the differences in (12) on day $d$ and $n_d$ is the total number of contact attempts made on that day:

$$\hat{B}_d = \sum_i \sum_t diff_{itd} / n_d \qquad (13)$$

$$se(\hat{B}_d) = s_d / \sqrt{n_d} \qquad (14)$$

5. We repeated Steps 1 through 4 above for $d = 8, \ldots, 84$ (NSFG quarters generally last 12 weeks), evaluating the mean difference and the standard error of that mean difference on each day $d$.

*Step 3: Evaluation of the Predictions.* We proceeded to evaluate trends in the daily mean differences from (13) for each method in each NSFG quarter. We also plotted the distributions *across* days of the mean differences in (13) (i.e., estimated biases) and estimates of the RMSE based on (13) and (14) (i.e., $\widehat{RMSE}_d = \sqrt{\hat{B}_d^2 + se(\hat{B}_d)^2}$ ) by quarter, for each of the methods. In these latter plots, we compared the performance of the methods early (days 7-30), midway through (days 31-60), and later (days 61-84) in each quarter.

## 3. Results

*3.1 Summaries of Final Response Propensity Models*

Table 1 presents summary statistics for the "final" discrete-time hazard models fitted in each of the five most recent quarters. Recall that these models generated our "final" benchmark predictions of response propensity in (9) for each case, using all available contact attempt data from each quarter.

**Table 1:** Model fit statistics for the "final" response propensity models fitted to all call-level data from each of the five most recent NSFG quarters.

|  | Quarter 16 | Quarter 17 | Quarter 18 | Quarter 19 | Quarter 20 |
|---|---|---|---|---|---|
| **Number of Calls** | 15,521 (3,431 interviews; 12,090 non-interviews) | 15,646 (3,668; 11,978) | 15,455 (3,431; 12,024) | 13,652 (3,426; 10,226) | 14,175 (3,373; 10,802) |
| **Nagelkerke Pseudo R-Squared** | 0.143 | 0.115 | 0.089 | 0.130 | 0.086 |
| **Hosmer-Lemeshow GOF test: p-value** | 0.094 | 0.547 | <0.01 | 0.033 | 0.432 |
| **AUC** | 0.711 | 0.682 | 0.661 | 0.690 | 0.654 |

Table 1 suggests that the "final" predictions of daily response propensity computed as benchmarks for each case (using all available data from each quarter) generally arose from discrete time hazard models with reasonable predictive power (Hosmer et al., 2013).

*3.2 Trends in Daily Differences across the Quarters*

Figures 2 and 3 present trends across the days of two of the five NSFG quarters (Quarters 17 and 19) in the mean differences between the daily predictions of response propensity and the final predictions of response propensity for each case (based on all data collected from the quarter); see (13). We specifically selected these two quarters for studying these trends in detail because comparisons of the performance of the approaches in aggregate revealed larger differences between the approaches than seen in the three other quarters (see Section 3.3). We also focus on the best-performing method (PWP) versus the standard method.

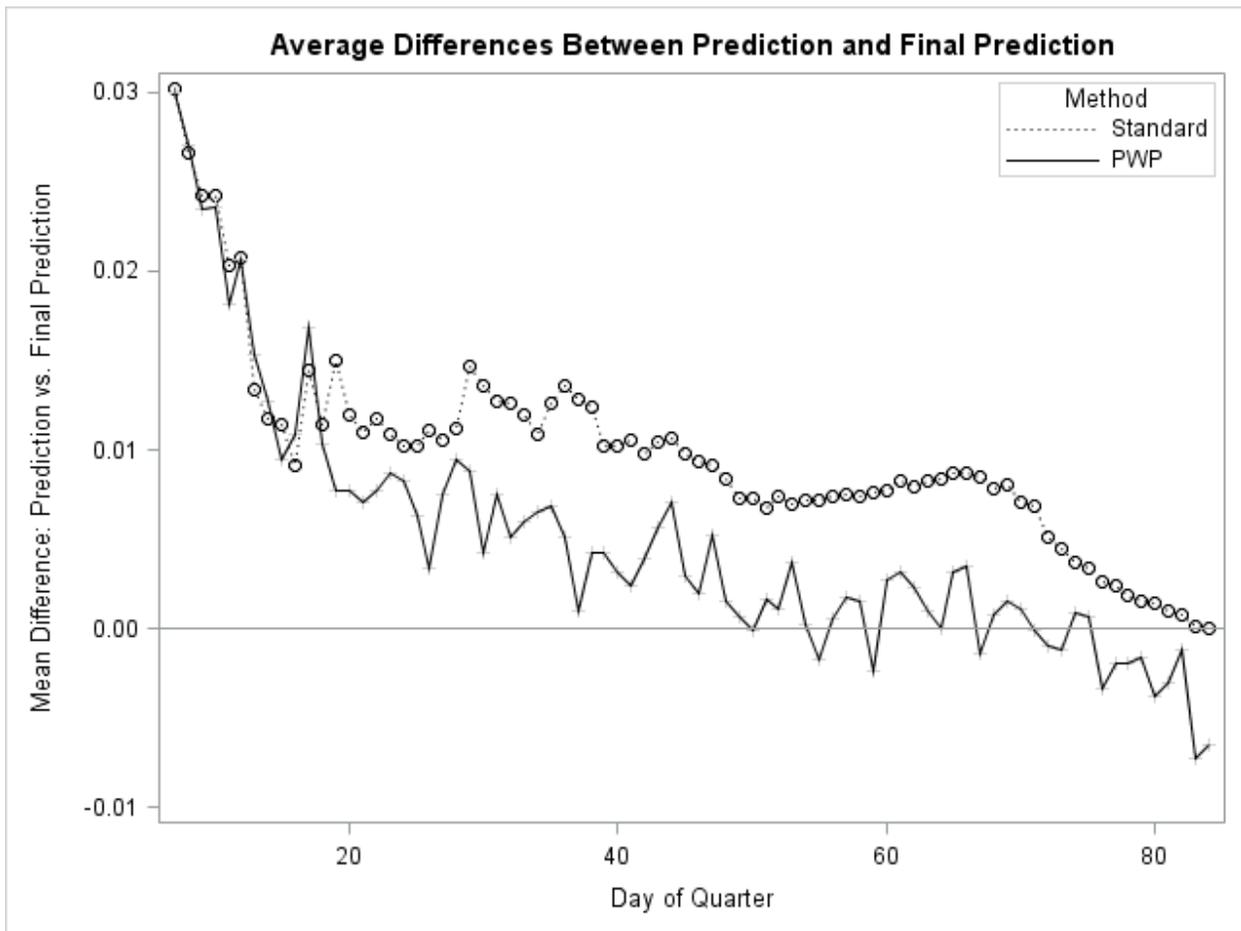

**Figure 2.** Trends in mean differences between daily predictions and final predictions across the 84 days in Quarter 17: Standard method vs. PWP method.

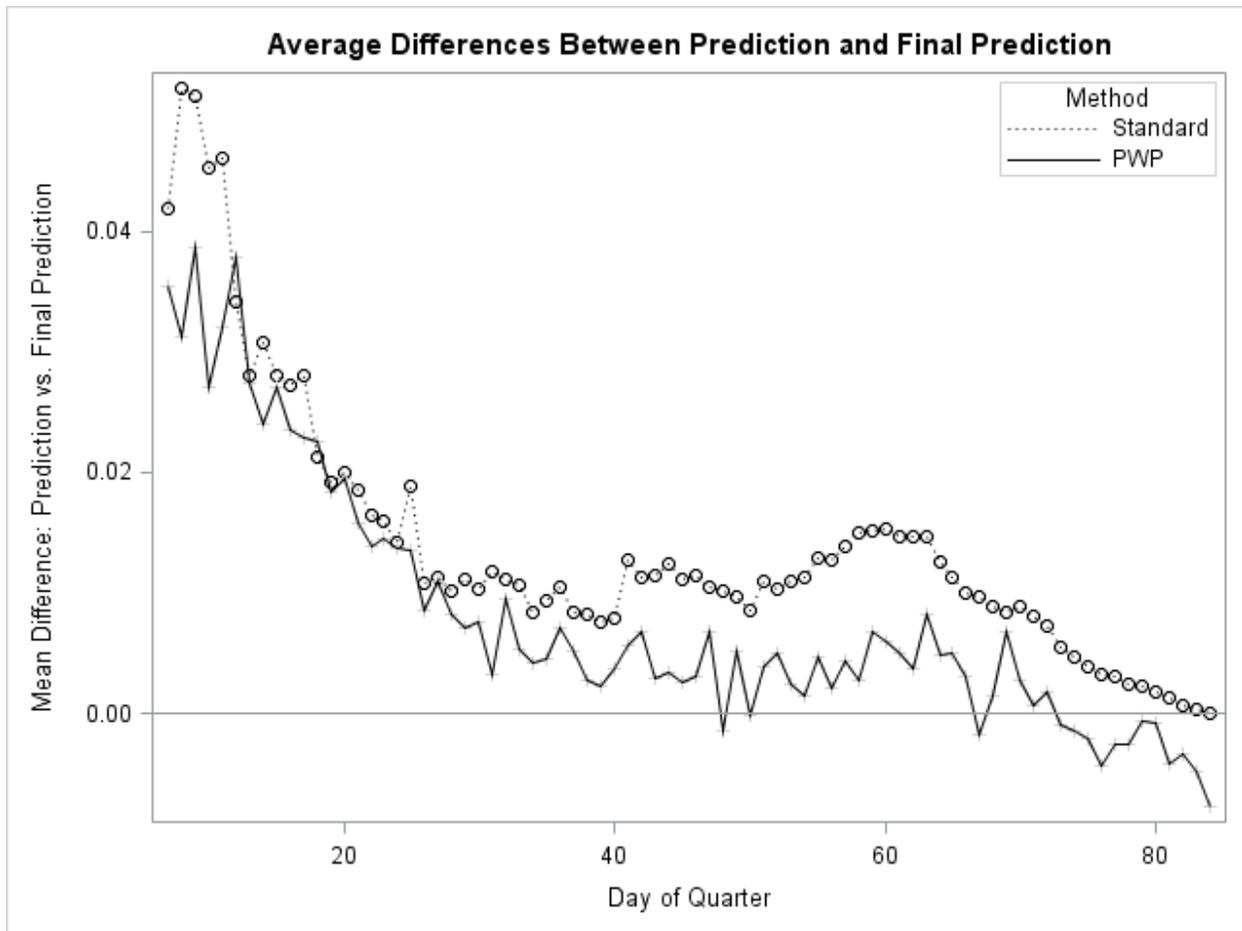

**Figure 3.** Trends in mean differences between daily predictions and final predictions across the 84 days in Quarter 19: Standard method vs. PWP method.

These figures demonstrate how predictions based on the PWP method tended to have noticeably lower mean differences compared to the standard method earlier in the data collection, and converged to zero on these differences more quickly. The largest differences also tended to emerge in the middle periods of the data collection, when interventions are often attempted (Wagner et al., 2012).

Figures A1 and A2 in the supplementary materials present the trends for all five methods in these same two quarters, including the standard errors of the estimated mean differences on each day. We note the relatively poor performance of the LAST method (relative to LASTZ) in Figure A1 and during later days in Figure A2, indicating the lack of stability in results based on this method across the quarters. We also find that the LIT method does not perform as consistently well as the PWP and LASTZ methods across the days of each quarter.

*3.3 Comparisons of Estimated Bias and RMSE of the Alternative Methods*

Figures 4 to 6 present the distributions of the mean differences in (13) based on days 7-30, 31-60, and 61-84, for each of the five methods by quarter. The consistent ability of the Bayesian approaches (with the exception of LAST) to shift the central tendencies of the estimated bias measures downward relative to the standard approach is apparent, especially during the "middle" periods of each quarter. We find that the PWP approach generally does quite well in this regard. We also note the general tendency of the predictions to approach the final predictions based on all data accumulated as the days in each quarter proceed, as would be expected.

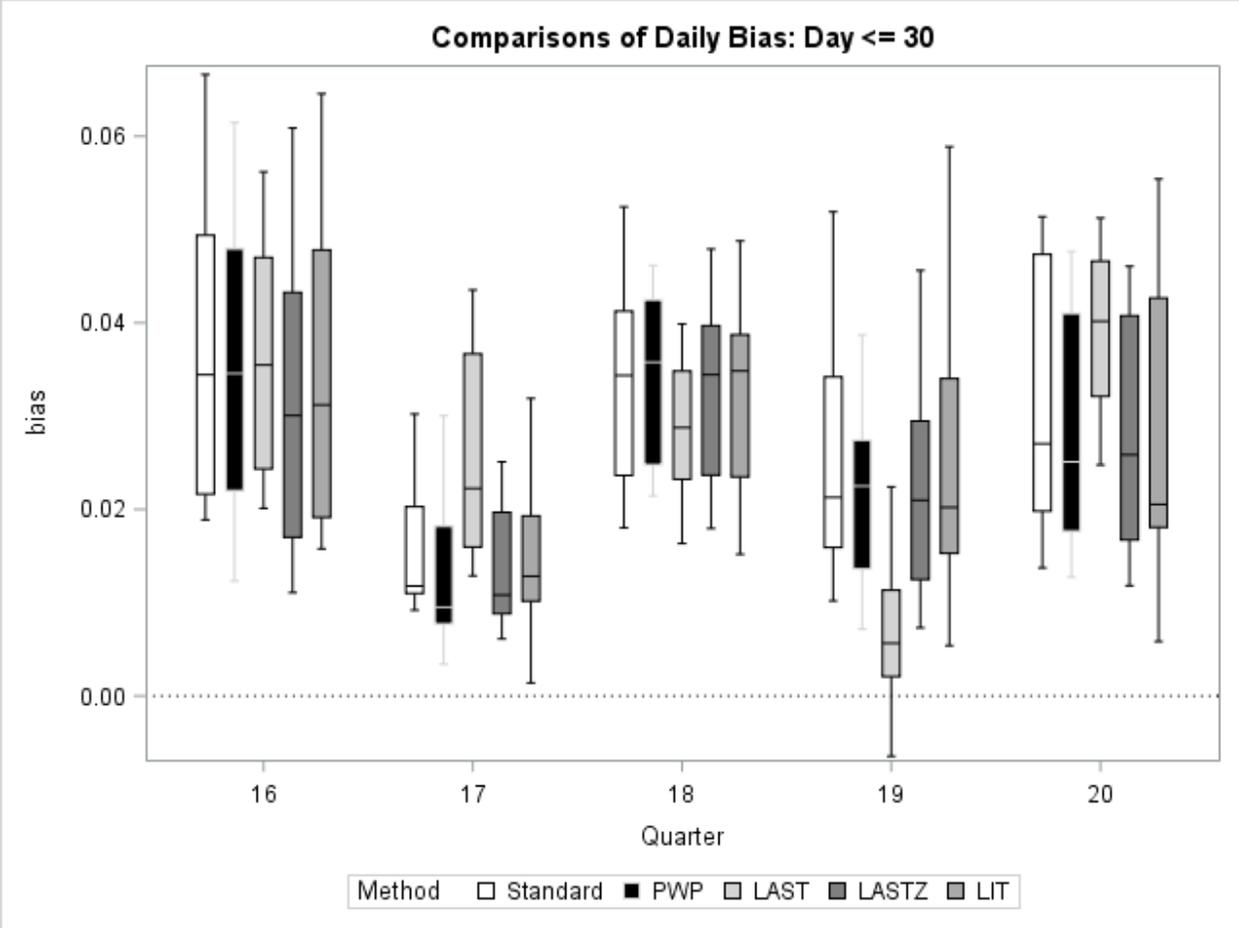

**Figure 4.** Distributions of mean differences (estimated bias) across days by prediction approach for each of the five quarters (days 7 – 30 only).

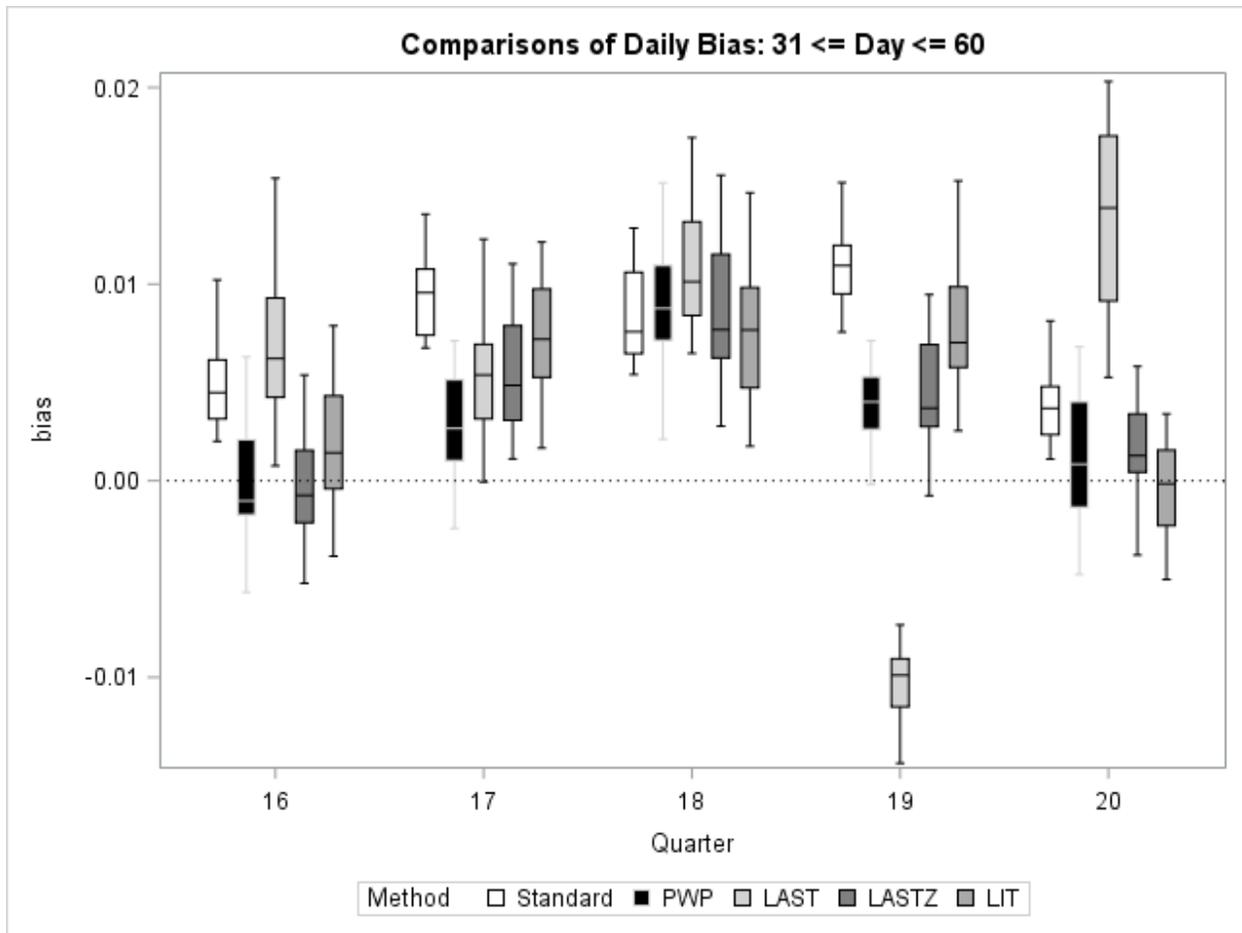

**Figure 5.** Distributions of mean differences (estimated bias) across days by prediction approach for each of the five quarters (days 31 – 60 only).

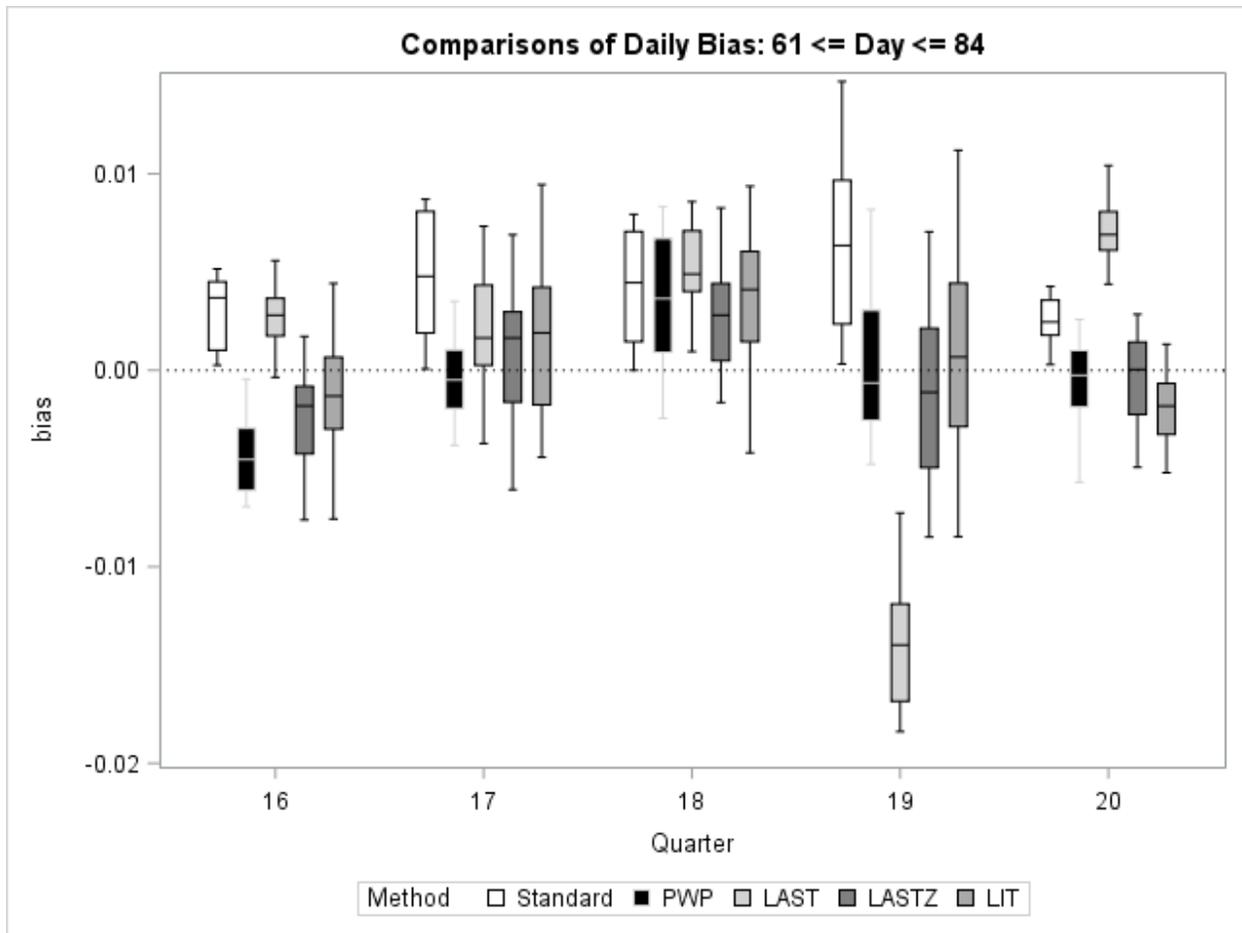

**Figure 6.** Distributions of mean differences (estimated bias) across days by prediction approach for each of the five quarters (days 61 – 84 only).

Figures 7 to 9 present the same comparisons in terms of the estimated RMSE of the mean differences. Similar patterns are evident here, again providing support for the Bayesian approaches when accounting for the estimated variances of the daily mean differences as well (especially in Quarters 17 and 19, as noted earlier). These plots also provide consistent evidence in favor of the approaches using historical data to formulate the priors (with the exception of the unstable performance of LAST), although the LIT approach can certainly be competitive.

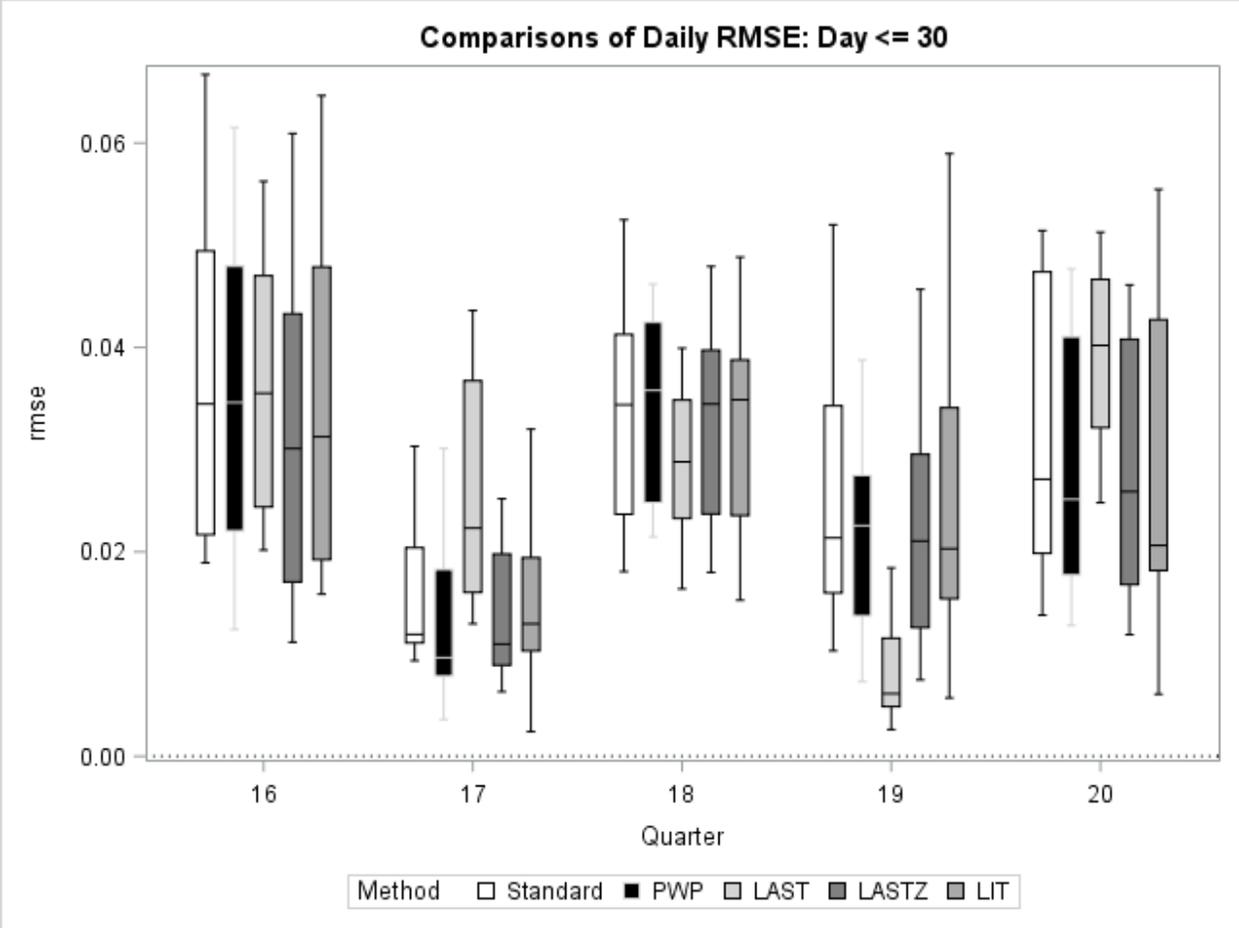

**Figure 7.** Distributions of estimated RMSE across days by prediction approach for each of the five quarters (days 7 – 30 only).

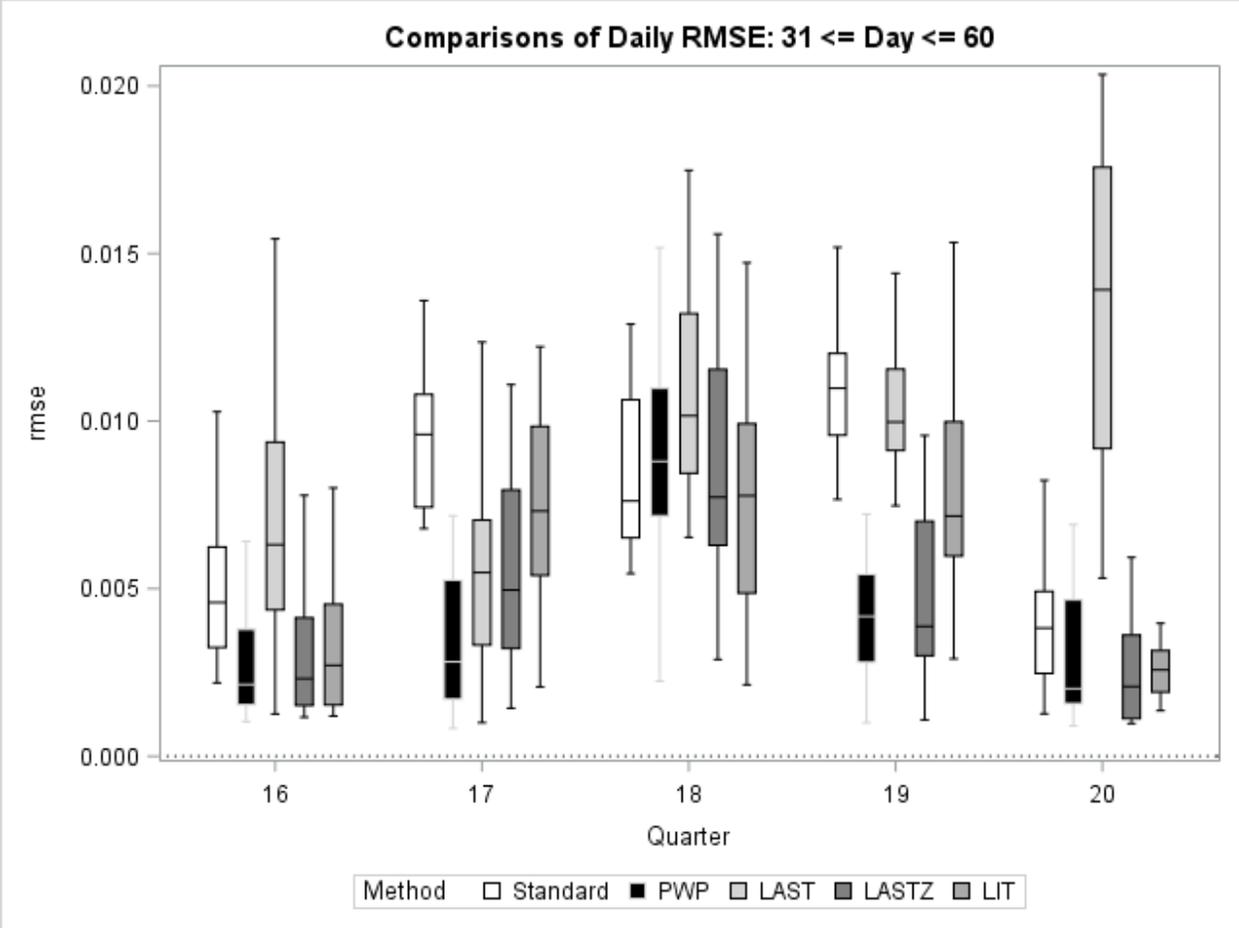

**Figure 8.** Distributions of estimated RMSE across days by prediction approach for each of the five quarters (days 31 – 60 only).

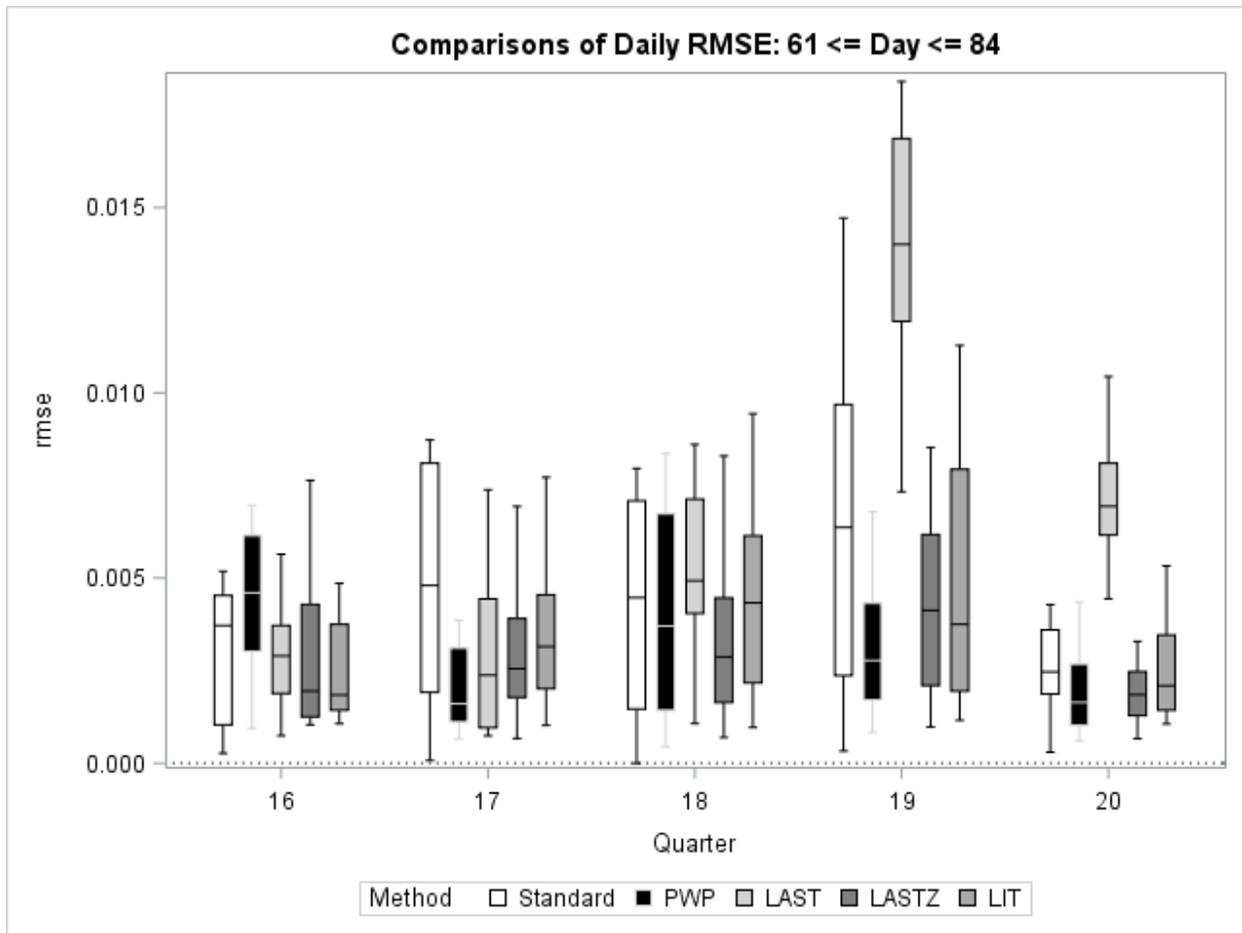

**Figure 9.** Distributions of estimated RMSE across days by prediction approach for each of the five quarters (days 61 – 84 only).

## 4. Discussion

*4.1 Summary of Findings*

We find evidence of improvements in both the bias and variance of predictions of daily response propensity (at the contact attempt level) in the NSFG by using Bayesian methods for estimating the underlying discrete-time logit models. This is especially true in the early-to-middle periods of a given NSFG data collection quarter, when survey managers often consider design modifications based on estimates of response propensity. When specifically considering the three different methods for deriving prior evidence on the model coefficients, we find general support for the PWP and LASTZ methods, where the PWP method is capable of leveraging a large amount of historical data, and the LASTZ method only requires evidence from a recent data collection. Notably, the method based on the prior literature (LIT) can also be competitive with these other methods leveraging historical data, suggesting that this is a reasonable approach to developing priors when historical data may not be available.

The NSFG was a useful case study in this context, given 1) its measurement of repeated, independent, national quarter-samples throughout the calendar year using the same design, and 2) its stability in response rates and response propensity models over time, which collectively make the derivation of priors based on the PWP method both possible and reasonable. If other surveys have not collected a similarly large amount of relevant historical data for deriving priors using the PWP method, or response propensities tend to vary over time or fluctuate seasonally, the LASTZ and LIT approaches would seem to be competitive alternatives. Our results suggest that new or one-off surveys interested in daily predictions of response propensity for responsive survey design could improve these predictions by deriving informative prior distributions for the coefficients defining their models using historical estimates from surveys with similar target populations and response propensity models.

One notable difficulty associated with the LIT approach involves variable selection for the response propensity models. For example, if a survey uses a variety of unique paradata to maximize the predictive power of a response propensity model, finding other studies using comparable models may be difficult, if not impossible. In addition, having very clear definitions of the paradata used in prior studies will be essential for confirming that the coefficients from the prior studies are relevant for the LIT approach. Even if one can find other studies presenting the coefficient(s) and standard errors for a specific predictor (which we could do for 46% of the coefficients in our response propensity model), those coefficients may come from a model that includes a *different* set of predictors, which could limit the utility of the prior information and thus the LIT approach. Replication of our approach in other contexts would be helpful for further assessing the value of the LIT approach, given these issues.

Our methods are easy to implement using statistical software enabling Bayesian computation. We used R and PROC MCMC in SAS (Version 9.4) in this study; our annotated code is available in the supplementary materials, and readers are encouraged to contact the authors for assistance with its use. The microdata used to fit the models described in this study are available from the National Center for Health Statistics. These microdata can be accessed via a restricted data user agreement; see https://www.cdc.gov/nchs/nsfg/nsfg_2017_2019_puf.htm and https://www.cdc.gov/rdc/ for details.

*4.2 Do Improved Predictions Lead to Better RSD Decisions?*

This paper has focused on the ability of Bayesian approaches leveraging prior information about the coefficients defining response propensity models to improve the model-based predictions of response propensity that are often used to make real-time design modification decisions in RSD. Improving the quality of these predictions is critical for being able to project the outcomes of survey data collections accurately (Wagner and Hubbard, 2014). How exactly these predictions are used to make design decisions during data collection was outside the scope of this paper. For example, active cases could be divided into broad subgroups based on their predicted response propensities (e.g., low/medium/high), where different design modifications may be applied to the different subgroups (e.g., using a more expensive data collection mode for cases with lower predicted response propensity). Whether the same active case would ultimately be assigned to different subgroups based on Bayesian vs. non-informative prediction methods is important to assess within the context of a specific survey and relative to specific interventions that will be assigned to those groups. This is an important next step in this line of research.

We also feel that experimental work is needed in a real survey to evaluate whether the improvements in predictions of response propensity demonstrated here ultimately lead to design modifications that improve survey efficiency, from the perspectives of cost and the mean squared error of survey estimates. A random half-sample of the sampled cases in a survey could be assigned to a "Bayesian" condition, where predictions of daily response propensity are computed using one of our approaches, and the decision rules planned in advance by a survey program employing RSD would apply these predictions when making design modifications. The other half-sample would have predictions computed in a "non-informative" manner, only relying on information from the current data collection. These two groups of sampled cases could then be compared in terms of the response rates, survey estimates, costs, and other key survey outcomes engendered by applying the design modifications, and one could examine whether these outcomes are improved by utilizing improved predictions of daily response propensity.

Careful consideration of the role of predictions of daily response propensity in formal decision frameworks for RSD is also important in thinking about these future evaluations. An important aspect of RSD is estimating future costs associated with different cases under different decisions about how they should be treated. Part of this cost estimation will rely on the propensity of that case to respond, which, based on the results of this work, can be estimated with greater accuracy when historical data or literature is available.

*4.3 What About Expert Elicitation?*

We did not consider one alternative that has also received attention in the health sciences literature: consultation with subject-matter experts to elicit their beliefs about the coefficients in these models (e.g., Boulet et al., 2019). We have also evaluated the potential of this approach to elicit the type of prior information analyzed in this study. We developed a simple questionnaire for 20 survey and data collection managers (e.g., interviewer supervisors) employed at the U.S. Census Bureau and the University of Michigan that collected information about expected call-level response rates in subgroups defined by the predictors in our response propensity model (see https://osf.io/3kxzb/), and recently completed this data collection.

After collecting data from these 20 managers, we translated the estimated call-level response rates into the coefficients of a logistic regression model, and then computed the mean and the variance of the coefficients across respondents to generate normal prior distributions for each of the coefficients of interest. Analyses similar to those performed in this paper have suggested that this prior elicitation approach competes well with the historical data and literature review approaches considered here; see Coffey et al. (2020).

*4.4 General Extensions*

We would welcome replications of our study in other survey design contexts, where the gains from the Bayesian approach may be larger if the prior distributions are more informative or the underlying models for selected design quantities have a stronger fit. Our study focused on improving the predictions of the daily response propensities that often feed into RSD, via the use of prior information about the coefficients defining the models used to compute these predictions. Following Schouten et al. (2018), one could certainly apply these techniques to improve predictions of other survey design quantities used to make decisions in an RSD framework. From a sampling perspective, improving predictions of the probability that a given sampled unit will contain an eligible person has the potential to improve sampling efficiency. Improving predictions of the costs associated with collecting data from a given sampled unit has the potential to inform sample allocation decisions when employing techniques that rely on cost estimates, such as optimal allocation of sample to different strata (Kish, 1965). Replications of this study in surveys with historical data readily available should be straightforward, but the literature review required for a different survey context may be more time-consuming.

Another extension of our work involves propensity model coefficients that change significantly across data collection periods (e.g., seasons) or possibly across days within a data collection period. In these settings, one may need different priors depending on how the coefficients vary over time; for example, when using the PWP approach, one may only want to combine estimates from the *first* quarter of several prior years, rather than *all* prior quarters, if there are strong seasonality effects. Prior distributions specific to a day (or a season) may be more informative than fixed prior distributions assuming that the coefficients are essentially stable over time. We did try using these types of dynamic priors in the NSFG context, but our results did not change substantially, as there was little evidence of seasonality in the response propensity coefficients. Alternatively, we could consider the variability of the point estimates themselves over prior quarters as part of a measure of the prior variance. Although not necessary for our context, these extensions could be useful for other surveys.